\documentclass[letterpaper]{article}

\usepackage{subfig}
\usepackage{gensymb}
\usepackage[group-separator={,}]{siunitx}
\usepackage{hyperref}
\usepackage[affil-it]{authblk}
\usepackage{amsmath}
\usepackage{graphicx}
\usepackage{geometry}
\usepackage[font={small,it}]{caption}

\begin{document}

% TITLE
\title{Informative and misinformative interactions in a school of fish}

% AUTHORS
\author[1,5]{Emanuele Crosato \thanks{emanuele.crosato@sydney.edu.au}}
\author[2,3]{Li Jiang} 
\author[3,4]{Valentin Lecheval}
\author[1]{Joseph T. Lizier}
\author[5]{X. Rosalind Wang}
\author[3]{Pierre Tichit}
\author[3]{Guy Theraulaz}
\author[1]{Mikhail Prokopenko}

% AFFILIATIONS
\affil[1]{Complex Systems Research Group and Centre for Complex Systems, Faculty of Engineering \& IT, The University of Sydney, Sydney, NSW 2006, Australia.}
\affil[2]{School of Systems Science, Beijing Normal University, Beijing, 100875, P. R. China.}
\affil[3]{Centre de Recherches sur la Cognition Animale, Centre de Biologie Int\'{e}grative (CBI), Centre National de la Recherche Scientifique (CNRS), Universit\'{e} Paul Sabatier (UPS), F-31062 Toulouse Cedex 9, France.}
\affil[4]{Groningen Institute for Evolutionary Life Sciences, University of Groningen, Centre
for Life Sciences, Nijenborgh 7, 9747AG Groningen, The Netherlands.}
\affil[5]{CSIRO Data61, PO Box 76, Epping, NSW 1710, Australia.}

% DATE
\date{\vspace{-8ex}}

\flushbottom
\maketitle
\thispagestyle{empty}

% ABSTRACT
\begin{abstract}
It is generally accepted that, when moving in groups, animals process information to coordinate their motion.
Recent studies have begun to apply rigorous methods based on Information Theory to quantify such distributed computation.
Following this perspective, we use transfer entropy to quantify dynamic information flows locally in space and time across a school of fish during directional changes around a circular tank, i.e. U-turns.
This analysis reveals peaks in information flows during collective U-turns and identifies two different flows: an informative flow (positive transfer entropy) based on fish that have already turned about fish that are turning, and a misinformative flow (negative transfer entropy) based on fish that have not turned yet about fish that are turning.
We also reveal that the information flows are related to relative position and alignment between fish, and identify spatial patterns of information and misinformation cascades.
This study offers several methodological contributions and we expect further application of these methodologies to reveal intricacies of self-organisation in other animal groups and active matter in general.
\end{abstract}

%%%%%%%%%%%%%%%%%%%%%%%%%%%%%%%%%%%%%%%%%%%%%%%%%%%%%%%%%%%%%%%%%%%%%%
%%%%%%%%%%%%%%%%%%%%%%%%%%%% INTRODUCTION %%%%%%%%%%%%%%%%%%%%%%%%%%%%%%%
%%%%%%%%%%%%%%%%%%%%%%%%%%%%%%%%%%%%%%%%%%%%%%%%%%%%%%%%%%%%%%%%%%%%%%

\section*{Introduction}

Collective motion is one of the most striking examples of aggregated coherent behaviour in animal groups, dynamically self-organising out of local interactions between individuals.
It is observed in different animal species, such as schools of fish~\cite{parrish2002self, sumpter2008information}, flocks of birds~\cite{lissaman1970formation, may1979flight, ballerini2008interaction, bialek2012statistical}, colonies of insects~\cite{buhl2006disorder, fourcassie2010ant, buhl2010group, attanasi2014collective, buhl2016} and herds of ungulates~\cite{ginelli2015intermittent}.
There is an emerging understanding that information plays a \emph{dynamic} role in such a coordination~\cite{sumpter2008information}, and that \emph{distributed} information processing is a specific mechanism that endows the group with collective computational capabilities~\cite{bonabeau1999swarm, couzin2009collective, albantakis2014evolution}.

Information transfer is of particular relevance for collective behaviour, where it has been observed that small perturbations cascade through an entire group in a wave-like manner~\cite{potts1984chorus, procaccini2011propagating, herbert2015initiation, attanasi2015emergence}, with these cascades conjectured to embody information transfer~\cite{sumpter2008information}.
This phenomenon is related to underlying causal interactions, and a common goal is to infer physical interaction rules directly from experimental data~\cite{katz2011inferring, gautrais2012deciphering, herbert2011inferring} and measure correlations within a collective.

Nagy et al.~\cite{nagy2010hierarchical} used a variety of correlation functions to measure directional dependencies between the velocities of pairs of pigeons flying in flocks of up to ten individuals, reconstructing the leadership network of the flock.
As has been shown later, this network does not correspond to the hierarchy between birds~\cite{nagy2013context}.
Information transfer has been extensively studied in flocks of starlings, by observing the propagation of direction changes across the flocks~\cite{cavagna2013diffusion, cavagna2013boundary, attanasi2014information}.
More recently, Rosenthal et al.~\cite{rosenthal2015revealing} attempted to determine a communication structure of a school of fish during its collective evasion manoeuvres manifested through cascades of behavioural change.
A functional mapping between sensory inputs and motor responses  was inferred by tracking fish position and body posture, and calculating visual fields.

Rather than consider \emph{semantic} or \emph{pragmatic} information, many contemporary studies employ rigorous information theoretic measures that quantify information as uncertainty reduction, following Shannon~\cite{cover91}, in order to deal with the stochastic, continuous and noisy nature of intrinsic information processing in natural systems~\cite{feldman2008organization}.
Distributed information processing is typically dissected into three primitive functions: the \emph{transmission}, \emph{storage} and \emph{modification of information}~\cite{langton1990computation}.
\emph{Information dynamics} is a recent framework characterising and measuring each of the primitives information-theoretically~\cite{liz14a,lizier2013book}.
In viewing the state update dynamics of a random process as an information processing event, this framework performs an \emph{information regression} in accounting for where the information to predict that state update can be found by an observer, first identifying predictive information from the past of the process as \emph{information storage}, then predictive information from other sources as \emph{information transfer} (including both pairwise transfer from single sources, and higher-order transfers due to multivariate effects).
The framework has been applied to modelling collective behaviour in several complex systems, such as Cellular Automata~\cite{lizier2008local, lizier2010information, lizier2012local}, Ising spin models~\cite{barnett2013information}, Genetic Regulatory Networks and other biological networks~\cite{lizier2011information, prokopenko2011relating, faes14a}, and neural information processing~\cite{gomez2014reduced, wibral2015bits}.

This study proposes a domain-independent, information-theoretic approach to detecting and quantifying individual-level dynamics of information transfer in animal groups using this framework. This approach is based on transfer entropy \cite{schreiber2000measuring}, an information-theoretic measure that quantifies the directed and time-asymmetric predictive effect of one random process on another.
We aim to characterize the dynamics of how information transfer is conducted in space and time within a \emph{biological} school of fish (\textit{Hemigrammus rhodostomus} or rummy-nose tetras, Figure \ref{fig:polarisation-fish}).

We stress that the predictive information transfer should be considered from the observer perspective, that is, it is the observer that gains (or loses) predictability about a fish motion, having observed another fish.
In other words, notwithstanding possible influences among the fish that could potentially be reflected in their information dynamics, our quantitative analysis focuses on the information flow within the school which is detectable by an external observer, captured by the transfer entropy.
This means that, whenever we quantify a predictive information flow from a source fish to a destination fish, we attribute the change of predictability (uncertainty) to a third party, be it another fish in the school, a predator approaching the school or an independent experimentalist.
Accordingly, this predictive information flow may or may not account for the causal information flow affecting the source and the destination~\cite{ay2008information, lizier2010differentiating} --- however it does typically indicate presence of causality, either within the considered pair or from some common cause.

We focus on collective direction changes, i.e. collective U-turns, during which the directional changes of individuals
progress in a rapid cascade, at the end of which a coherent motion is re-established within the school.
Sets of different U-turns are comparable across experiments under the same conditions, permitting a statistically significant analysis involving an entire set of U-turns.

By looking at the \emph{pointwise} or \emph{local} values of transfer entropy over time, rather than at its average values, we are not only able to detect information transfer, but also to observe its dynamics over time and across the school.
We demonstrate that information is indeed constantly flowing within the school, and identify the source-destination lag where predictive information flow is maximised (which has an interpretation as an observer-detectable reaction time to other fish).
The information flow is observed to peak during collective directional changes, where there is a typical ``cascade'' of predictive gains and losses to be made by observers of these pairwise information interactions.
Specifically, we identify two distinct predictive information flows: (i) an ``informative'' flow, characterised by positive local values of transfer entropy, based on fish that have already changed direction about fish that are turning, and (ii) a ``misinformative'' flow, characterised by negative local values of transfer entropy, based on fish that have not changed direction yet about the fish that are turning.
Finally, we identify spatial patterns coupled with the temporal transfer entropy, which we call spatio-informational motifs.
These motifs reveal spatial dependencies between the source of information and its destination, which shape the directed pairwise interactions underlying the informative and misinformative flows.
The strong distinction revealed by our quantitative analysis between informative and misinformative flows is expected to have an impact on modelling and understanding the dynamics of collective animal motion.

%%%%%%%%%%%%%%%%%%%%%%%%%%%%%%%%%%%%%%%%%%%%%%%%%%%%%%%%%%%%%%%%%%%%%%
%%%%%%%%%%%%%%%%%%%%%%%%%%%%%% BACKGROUND %%%%%%%%%%%%%%%%%%%%%%%%%%%%%%%
%%%%%%%%%%%%%%%%%%%%%%%%%%%%%%%%%%%%%%%%%%%%%%%%%%%%%%%%%%%%%%%%%%%%%%

\section*{Information-theoretic measures for collective motion}

The study of Wang et al.~\cite{wang2012quantifying} introduced the use of transfer entropy to investigations of collective motion.
This work quantitatively verified the hypothesis that information cascades within an (artificial) swarm can be spatiotemporally revealed by \emph{conditional transfer entropy}~\cite{lizier2008local, lizier2010information} and thus correspond to communications, while the collective memory can be captured by \emph{active information storage}~\cite{lizier2012local}.

Richardson et al.~\cite{richardson2013dynamical} applied related variants of conditional mutual information, a measure of non-linear dependence between two random variables, to identify dynamical coupling between the trajectories of foraging meerkats.
Transfer entropy has also been used to study the response of schools of zebrafish to a robotic replica of the animal~\cite{butail2014information, ladu2015acute}, and to infer leadership in pairs of bats~\cite{orange2015transfer} and simulated zebrafish~\cite{butail2016model}.
Lord et al.~\cite{lord2016inference} also posed the question of identifying individual animals which are directly interacting with other individuals, in a swarm of insects (\emph{Chironomus riparius}).
Their approach used conditional mutual information (called ``causation entropy'' although it does not directly measure causality \cite{lizier2010differentiating}), inferring ``information flows'' within the swarm over moving windows of time.

Unlike the study of Wang et al.~\cite{wang2012quantifying}, the above studies quantified average dependencies over time rather than local dependencies at specific time points; for example, leadership relationships in general rather than their (local) dynamics over time.
Local versions of transfer entropy and active information storage have been used to measure pairwise correlations in a ``swarm'' of soldier crabs, finding that decision-making is affected by the group size~\cite{tomaru2016information}.
Statistical significance was not reported, presumably due to a small sample size.
Similar techniques were used to construct interaction networks within teams of simulated RoboCup agents~\cite{cliff2017quantifying}.

In this study we focus on local (or pointwise) transfer entropy~\cite{schreiber2000measuring,lizier2008local,liz14b} for specific samples of time-series processes of fish motion, which allows us to reconstruct the dynamics of information flows over time.
Local transfer entropy~\cite{lizier2008local}, captures information flow from the realisation of a \textit{source} variable $Y$ to a \textit{destination} variable $X$ at time $n$.
As described in Methods, local transfer entropy is defined as the information provided by the source $\mathbf{y_{n-v}} = \{y_{n-v}, y_{n-v-1}, \ldots, y_{n-v-l+1}\}$, where $v$ is a time delay and $l$ is the history length, about the destination $x_n$ in the context of the past of the destination $\mathbf{x_{n-1}}=\{x_{n-1}, x_{n-2}, \ldots, x_{n-k}\}$, with a history length $k$:
\begin{equation}
t_{y\to x}(n,v) = \log_2\frac{ p( x_{n} | \mathbf{x_{n-1}} , \mathbf{y_{n-v}} ) }{ p( x_{n} | \mathbf{x_{n-1}} ) }.
\label{eq:teInBackground}
\end{equation}
Importantly, local values of transfer entropy can be negative, while the average transfer entropy is non-negative.
Negative values of the local transfer entropies indicate that the source is \textit{misinformative} about the next state of the destination (i.e. it increases uncertainty).
Previous studies that used average measures over sliding time windows in order to investigate how information transfer varies over time~\cite{richardson2013dynamical, lord2016inference} cannot detect misinformation because they measure average but not local values.

As an observational measure, transfer entropy does not measure causal effect of the source on the target; this can only be established using interventional measures \cite{ay2008information, lizier2010differentiating, chich12a, smirnov2013spurious}. Rather, transfer entropy measures the predictive information gained from a source variable about the state transition in a target, which may be viewed as \emph{information transfer} when measured on an underlying causal interaction \cite{lizier2010differentiating}.
It should be noted that while some researchers may be initially more interested in causality, the concept of information transfer reveals much about the dynamics that causal effect does not \cite{lizier2010differentiating}, in particular being associated with emergent local structure in dynamics in complex systems \cite{lizier2008local,wang2012quantifying} and with changes in behaviour, state or regime \cite{boedecker2012information,barnett2013information}, as well as revealing the misinformative interactions described above. As a particular example, local transfer entropy spatiotemporally highlights emergent glider entities in cellular automata \cite{lizier2008local}, which are analogues of cascading turning waves in swarms (also highlighted by transfer entropy \cite{wang2012quantifying}), while local measures of causality do not differentiate these from the background dynamics \cite{lizier2010differentiating}.

It is well known that the internal dynamics within a school of fish depends on the number of fish. For example, for schools of minnows (\textit{Phoxinus phoxinus}), two fish schools are qualitatively different from schools containing three or more --- however, the effects seem to level off by the time the school reaches a size of six individuals~\cite{partridge1980effect}.
Collective behaviour, as well as a stereotypical ``phase transition'', when an increase in density leads to the onset of directional collective motion, have also been detected in small groups of six glass prawns (\textit{Paratya australiensis})~\cite{mann2013multi}.
Furthermore, at such intermediate group sizes, it has been observed that multiple fish interactions could often be faithfully factorised into pair interactions in one particular species of fish~\cite{gautrais2012deciphering}.

In our study we investigated information transfer within a school of fish during specific collective direction changes, i.e., U-turns, in which the school collectively reverses its direction.
Groups of five fish were placed in a ring-shaped tank (Figure \ref{fig:polarisation-tank}), a design conceived to constrain fish swimming circularly, with the possibility of undergoing U-turns spontaneously, without any obstacles or external factors.
A total of 455 U-turns have been observed during 10 trials of one hour duration each.
We computed local transfer entropy between each (directed) pair of fish from time series obtained from fish heading.
Specifically, the destination process $X$ was defined as the directional change of the destination fish, while the source process $Y$ was defined as the relative heading of the destination fish with respect to the source fish (see Methods).
This allowed us to capture the influence of the source-destination fish alignment on the directional changes of the destination.
Such influence is usually delayed in time and we estimated the optimal delay (maximizing $\langle t_{y\to x}(n,v) \rangle_n$ \cite{wib13a}, see Methods) at $v=6$, corresponding to $0.12$ seconds.

%%%%%%%%%%%%%%%%%%%%%%%%%%%%%%%%%%%%%%%%%%%%%%%%%%%%%%%%%%%%%%%%%%%%%%
%%%%%%%%%%%%%%%%%%%%%%%%%%%%%% RESULTS %%%%%%%%%%%%%%%%%%%%%%%%%%%%%%%%%%
%%%%%%%%%%%%%%%%%%%%%%%%%%%%%%%%%%%%%%%%%%%%%%%%%%%%%%%%%%%%%%%%%%%%%%

\section*{Results}

%%%%%%%%%%%%%%%%%%%%%%%%%%
%%% Information flows during U-turns %%%
%%%%%%%%%%%%%%%%%%%%%%%%%%

\subsection*{Information flows during U-turns}

\begin{figure}[t]
\centering
\begin{minipage}{0.19\textwidth}
\centering
\subfloat[]{\label{fig:polarisation-fish}\includegraphics[width=\textwidth]{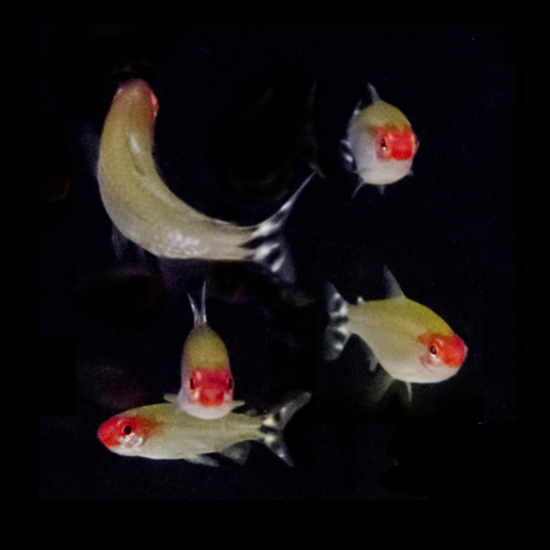}}\\\vspace{-3mm}
\subfloat[]{\label{fig:polarisation-tank}\includegraphics[width=\textwidth]{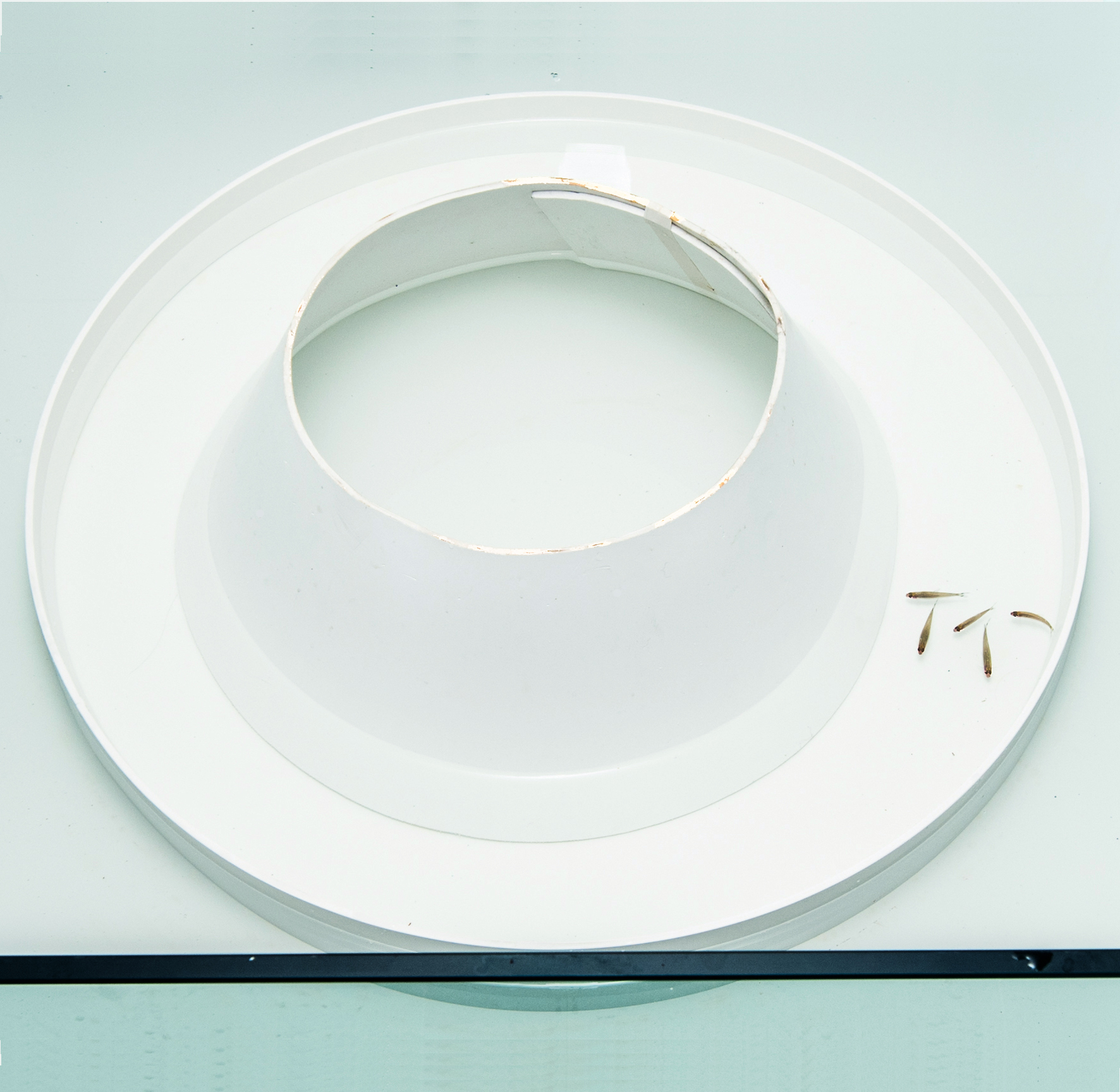}}
\end{minipage}
\hspace{5mm}
\begin{minipage}{0.6\textwidth}
\centering
\subfloat[]{\label{fig:polarisation-te}\includegraphics[width=\textwidth]{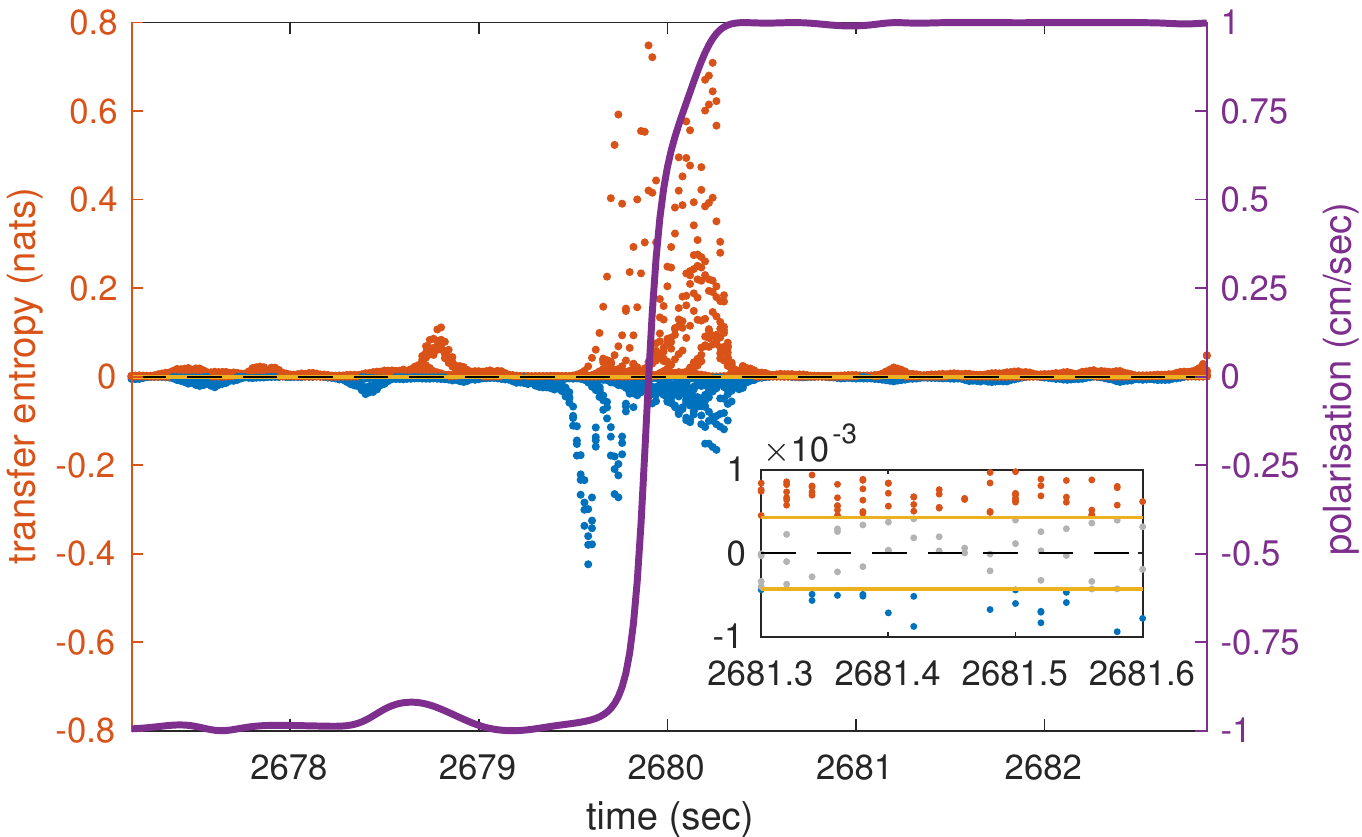}}
\end{minipage}
\caption{
Transfer entropy within the school during a U-turn.
Figure \ref{fig:polarisation-fish} is a photo of a spontaneous U-turn initiated by a single fish in a group of five \textit{Hemigrammus rhodostomus} fish.
Figure \ref{fig:polarisation-tank} shows the experimental ring-shaped tank.
Figure \ref{fig:polarisation-te} plots the school's polarisation during a U-turn and the detected transfer entropy over a time interval of approximately 6 seconds.
The purple line represents the school's polarisation, while dots represent local values of transfer entropy between all directed pairs of fish: red dots represent positive transfer entropy and blue dots represent negative transfer entropy.
Time is discretised in steps of length 0.02 seconds and for each time step 20 points of these local measures are plotted, for the 20 directed pairs formed out of 5 fish.
The yellow lines in the inset are the thresholds for considering a value of transfer entropy statistically different from zero ($p<0.05$ before false discovery rate correction, see Methods).
Grey dots between these lines represent values that are not statistically different from zero.
}
\label{fig:polarisation}
\end{figure}

In order to represent the school's orientation around the tank, we define its polarisation so that it is positive when the school is swimming clockwise and negative when it is swimming anti-clockwise (see Methods).
The better the school's average heading is aligned with an ideal circular trajectory around the tank, the higher is the intensity of the polarisation.
When the school is facing one of the tank's walls, for example in the middle of a U-turn, the polarisation is zero, and the polarisation flips sign during U-turns.
Polarisation allows us to map local values of transfer entropy onto the progression of the collective U-turns.

The analyses of transfer entropy over time reveal that the measure clearly diverges from its baseline in the vicinity of U-turns, as shown in the representative U-turn in Figure \ref{fig:polarisation-te} (Supplementary Figure S1 shows a longer time interval during which several U-turns can be observed).
The figure shows that during regular circular motion, when the school's polarisation is highly pronounced, transfer entropy is low.
As the polarisation approaches zero the intensity of transfer entropy grows, peaking near the middle of a U-turn, when polarisation switches its sign.

We clarify that the aim here is \emph{not} to establish transfer entropy as an alternative to polarisation for detecting turn; rather, our aim is to use polarisation to describe the overall progression of the collective U-turns and then to use transfer entropy to investigate the underlying information flows in the dynamics of such turns.
Indeed, transfer entropy is found to be statistically different from zero at many points outside of the U-turns (see Supplementary Figure S1), although the largest values and most concentrated regions of these are during the U-turns.
This indicates that information transfer occurs even when fish school together without changing direction; we know that the fish are not executing precisely uniform motion during these in-between periods, and so interpret these small amounts of information transfer as sufficiently underpinning the dynamics of the group maintaining its collective heading.

We also see in Figure \ref{fig:polarisation-te} that both positive and negative values of transfer entropy are detected.
In order to understand the role of the positive and negative information flows during collective motion, in the next section we show the dynamics of transfer entropy for individual pairwise interactions.

%%%%%%%%%%%%%%%%%%%%%%%%%%%%
%%% Informative and misinformative flows %%%
%%%%%%%%%%%%%%%%%%%%%%%%%%%%

\subsection*{Informative and misinformative flows}

\begin{figure}[t!]
\centering
\subfloat[]{\label{fig:trajectories-only}\includegraphics[width=0.32\textwidth]{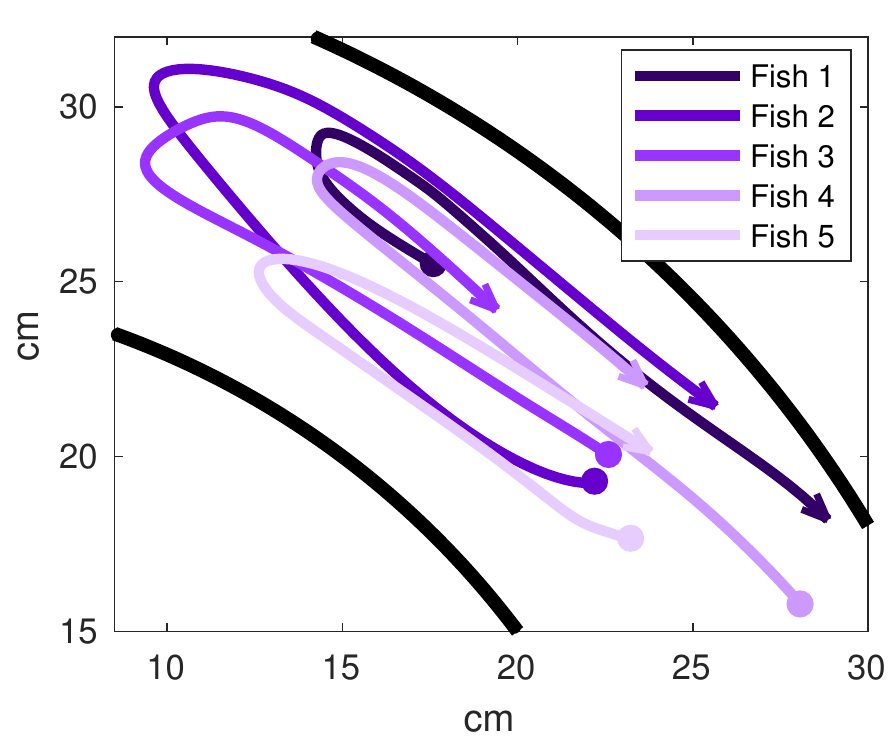}}\hspace{0.01\textwidth}
\subfloat[]{\label{fig:trajectories-in}\includegraphics[width=0.32\textwidth]{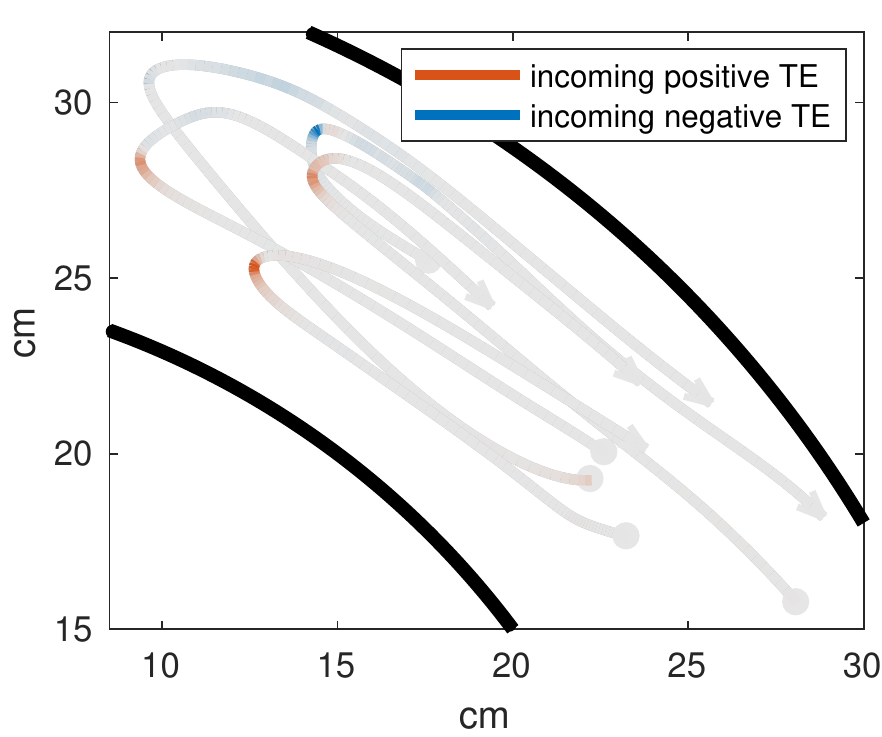}}\hspace{0.01\textwidth}
\subfloat[]{\label{fig:trajectories-out}\includegraphics[width=0.32\textwidth]{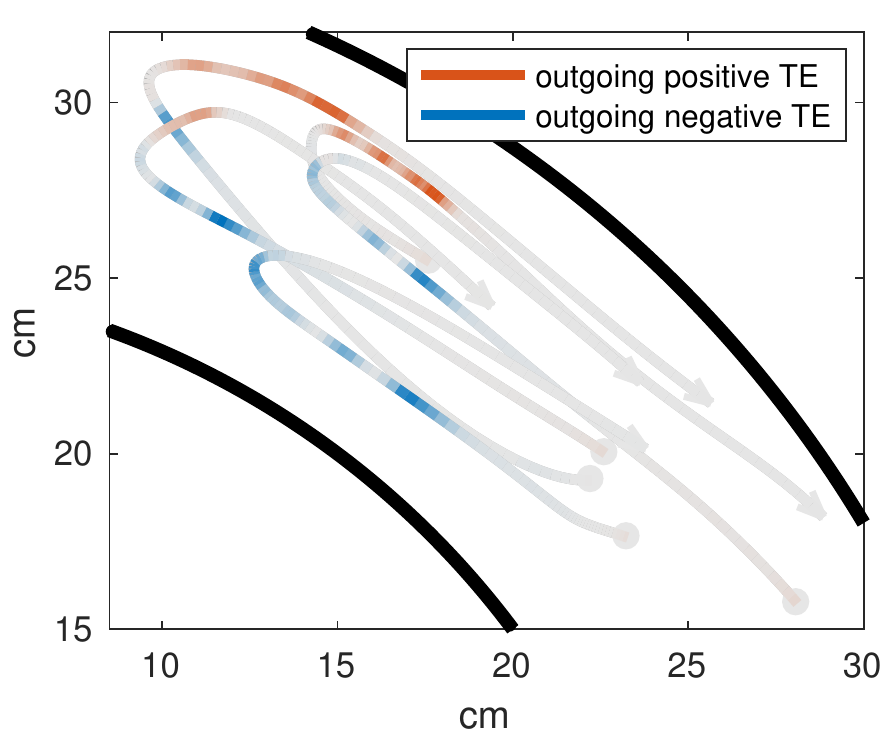}}\\
\subfloat[]{\label{fig:polarisations-only}\includegraphics[width=0.32\textwidth]{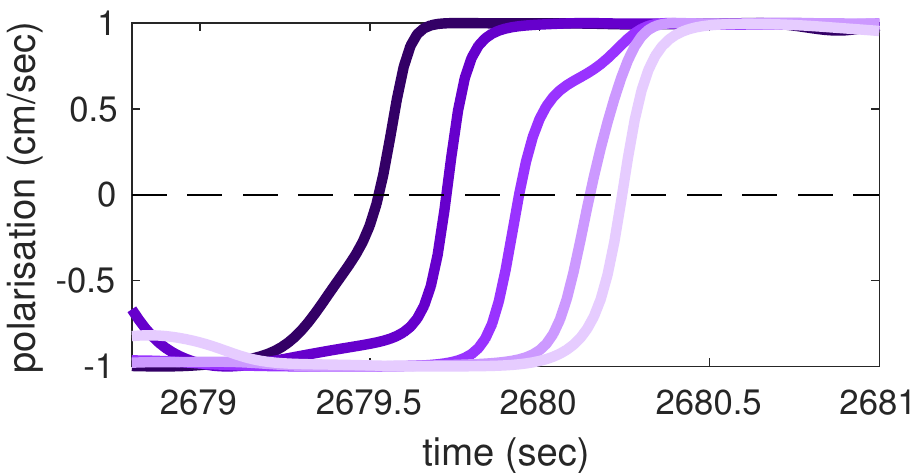}}\hspace{0.01\textwidth}
\subfloat[]{\label{fig:polarisations-in}\includegraphics[width=0.32\textwidth]{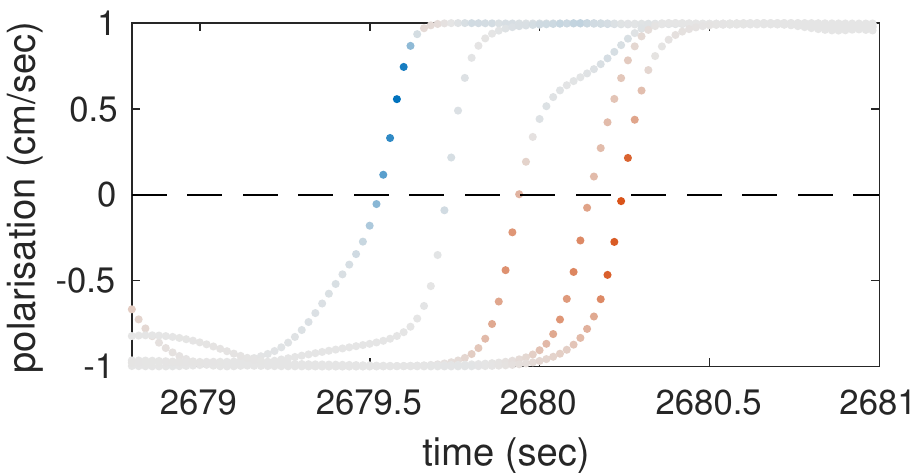}}\hspace{0.01\textwidth}
\subfloat[]{\label{fig:polarisations-out}\includegraphics[width=0.32\textwidth]{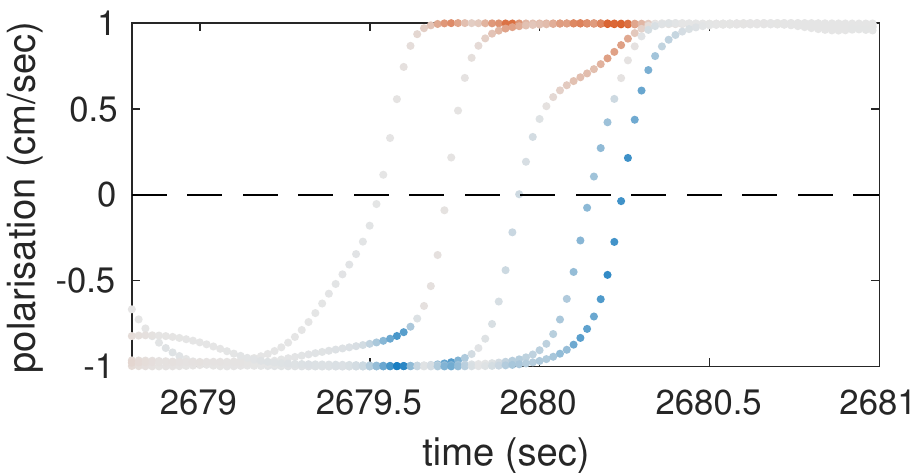}}
\caption{
Positive and negative information flows during a U-turn.
Figure \ref{fig:trajectories-only} shows the trajectories of the five fish during the U-turn shown in Figure \ref{fig:polarisation}.
The two black lines are the inner and the outer walls of the tank, and each of the five trajectories coloured in different shades of purple correspond to a different fish: from darkest purple for the first fish turning (Fish 1), to the lightest purple for the last (Fish 5).
The total time interval is approximately 2 seconds, during which all fish turn from swimming anti-clockwise to clockwise.
Figure \ref{fig:polarisations-only} depicts the polarisations of the five fish, showing the temporal sequence of fish turns.
Figure \ref{fig:trajectories-in} shows the fish trajectories again, but this time indicates the value of the \textit{incoming} local transfer entropy to each fish as a destination, averaged over the other four fish as sources. 
The colour of each trajectory changes as the fish turn:~strong red indicates intense positive transfer entropy; strong blue indicates intense negative transfer entropy; intermediate grey indicates that transfer entropy is close to zero.
Figure \ref{fig:polarisations-in} is obtained analogously to Figure \ref{fig:trajectories-in}, but the polarisations of the individual fish are shown rather than their trajectories.
Figures \ref{fig:trajectories-out} and \ref{fig:polarisations-out} mirror Figures \ref{fig:trajectories-in} and \ref{fig:polarisations-in}, but where the direction of the transfer entropy has been inverted: the colour of each trajectory or polarisation now indicates the value of the \textit{outgoing} local transfer entropy from each fish as a source, averaged over the other four fish as destinations. 
}
\label{fig:trajectories}
\end{figure}

\begin{figure}[t]
\centering
\subfloat[]{\label{fig:single-pol-in}\includegraphics[width=0.4\textwidth]{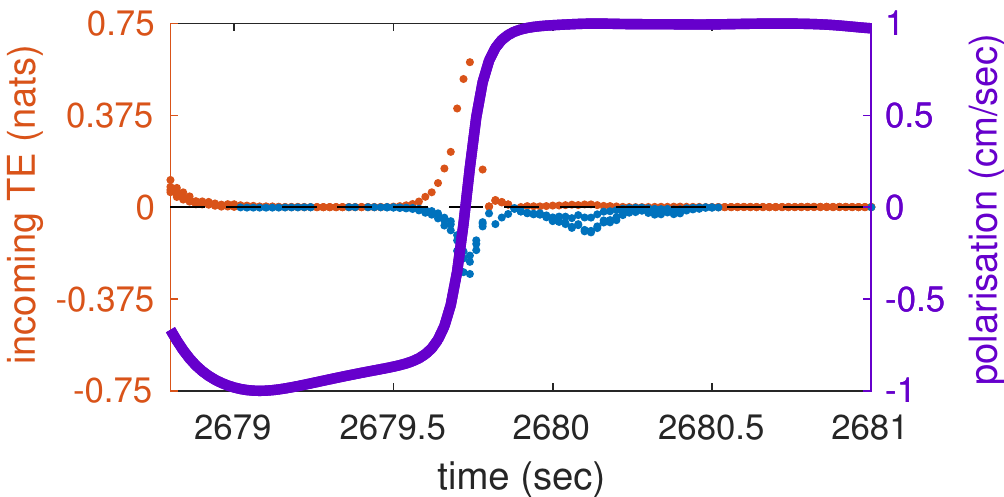}}\hspace{0.03\textwidth}
\subfloat[]{\label{fig:single-pol-out}\includegraphics[width=0.4\textwidth]{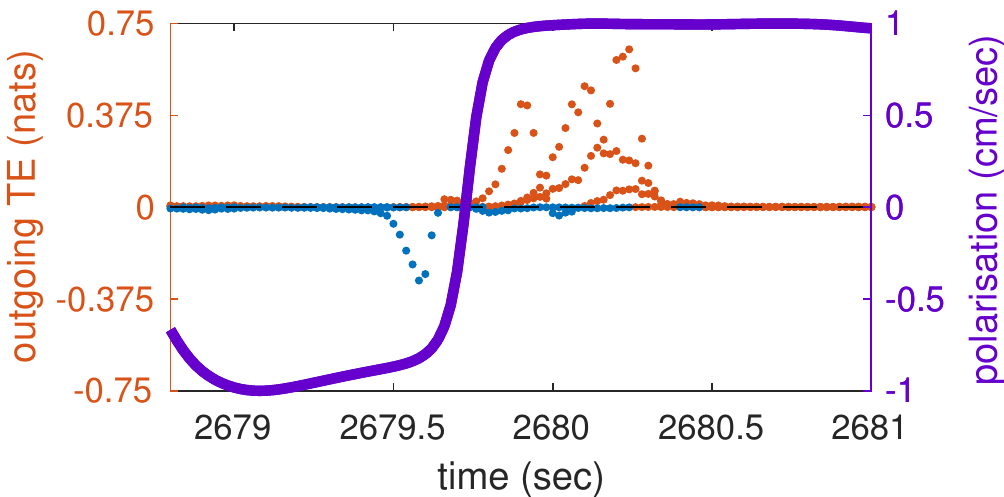}}\\
\subfloat[]{\label{fig:sequence-in}\includegraphics[width=0.37\textwidth]{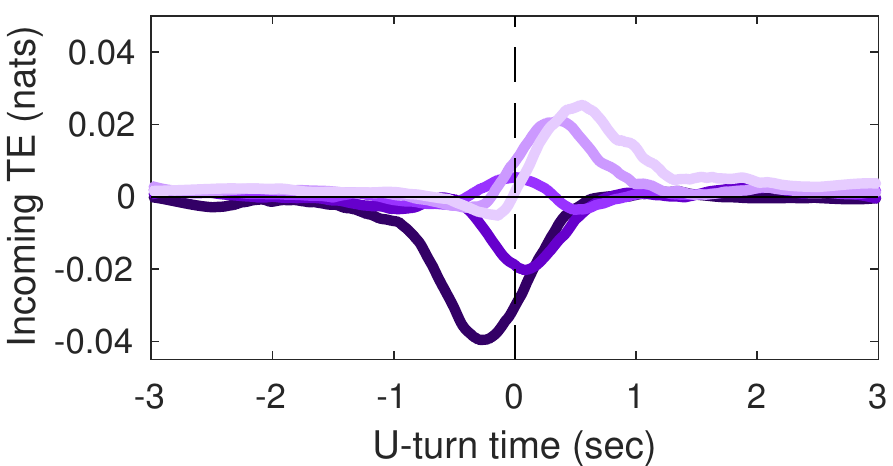}}
\hspace{0.03\textwidth}
\subfloat[]{\label{fig:sequence-out}\includegraphics[width=0.37\textwidth]{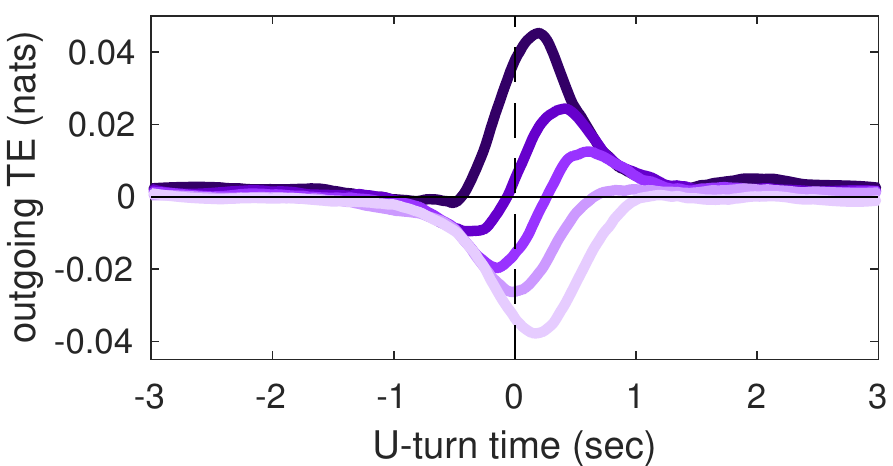}}
\caption{
Figure \ref{fig:single-pol-in} shows the polarisation of the \emph{second} fish turning, together with the incoming transfer entropy to that fish as the destination, with the other four fish as the sources:~red dots represent positive values and blue dots represent negative values.
Figure \ref{fig:single-pol-out} mirrors Figure \ref{fig:single-pol-in}, but with the outgoing transfer entropy from that fish as the source, and the other four fish as destinations.
In Figure \ref{fig:sequence-in}, each purple line corresponds to a fish, with the shade again reflecting the order in which the fish turn (darkest for first fish to turn, and lightest for the last).
Now however (in Figure \ref{fig:sequence-in}), rather than corresponding to a single U-turn event, the incoming local transfer entropy (to each fish as a destination, averaged over the other four fish as sources) is averaged over all 455 observed U-turns and is shown as a function of time.
The horizontal axis is the relative time of the U-turns, with zero being the time when the average polarisation of the school changes sign.
Figure \ref{fig:sequence-out} mirrors Figure \ref{fig:sequence-in}, but where the direction of the transfer entropy has been inverted (showing outgoing transfer from each fish in turning order).
}
\label{fig:cascade}
\end{figure}

Our analysis revealed a clear relationship between positive and negative values of transfer entropy and the sequence of individual fish turning, which is illustrated in Figure \ref{fig:trajectories}.
Figure \ref{fig:trajectories-only} shows the trajectories of individual fish during the same U-turn depicted in Figure \ref{fig:polarisation}.
These trajectories are retraced in Figure \ref{fig:polarisations-only} in terms of polarisation of each fish.
It is quite clear that there is a well-defined sequence of individual U-turns during the collective U-turn.
Moreover, Figure \ref{fig:trajectories} shows how the transfer entropy maps onto the fish trajectories, both from the fish whose trajectory is traced as a source to the other four fish~---~i.e. \textit{outgoing} transfer entropy~---~and, vice versa, from the other four fish to the traced one as a destination~---~i.e. \textit{incoming} transfer entropy.

The incoming transfer entropy clearly peaks during the destination fish's individual turns and its local values averaged over all sources go from negative, for the first (destination) fish that turns, to positive for the last fish turning (Figures \ref{fig:trajectories-in} and \ref{fig:polarisations-in}).
In the opposite direction, the outgoing transfer entropy (averaged over all destinations) displays negative peaks only before the source fish has turned, and positive peaks only afterwards (Figures \ref{fig:trajectories-out} and \ref{fig:polarisations-out}).
Figure \ref{fig:trajectories} suggests that predictive information transfer intensifies only when a destination fish is turning, with this transfer being informative based on source fish that have already turned and misinformative based on source fish that have not turned yet.

This phenomenon can be observed very clearly in Figures \ref{fig:single-pol-in} and \ref{fig:single-pol-out}, which show the transfer entropy in both directions for a single fish (the second fish turning in Figures \ref{fig:polarisation} and \ref{fig:trajectories}).
One positive peak of incoming transfer entropy (indicating informative flow) and three negative ones (misinformative flows) are detected when this fish, as a destination, is undergoing the U-turn (Figure \ref{fig:single-pol-in}).
No other peaks are detected for this fish as a destination.
On the other hand, one negative peak of outgoing transfer entropy is detected before the fish, this time as a source, has turned, and three positive peaks are detected after the fish has turned (Figure \ref{fig:single-pol-out}).
These four peaks occur respectively when the first, the third, the fourth and the fifth fish undergo the U-turn, as is evident by comparing Figures \ref{fig:single-pol-out} and \ref{fig:polarisations-only}.
A movie of the fish undergoing this specific U-turn is provided in Supplementary Video S1, while a detailed reconstruction of the U-turn, showing the dynamics of transfer entropy over time for each directed pair of fish, is provided in Supplementary Video S2.

In order to demonstrate that the phenomenon described here holds for U-turns in general, and not only for the representative one shown in Figure \ref{fig:trajectories}, we performed an aggregated analysis of all 455 U-turns observed during the experiment.
Since the order in which fish turn is not the same in every U-turn, in this analysis, we refer not to single fish as individuals, but rather to fish in the order in which they turn.
Thus, when we refer, for instance, to ``the first fish that turns'', we may be pointing to a different fish at each U-turn.

The aggregated results are presented in Figures \ref{fig:sequence-in} and \ref{fig:sequence-out}.
Figure \ref{fig:sequence-in} shows that incoming transfer entropy peaks for each fish in turning order and gradually grows, from a minimum negative peak corresponding to the first fish turning, to a maximum positive peak corresponding to the last fish turning.
Vice versa, Figure \ref{fig:sequence-out} shows that outgoing transfer entropy peaks only positively for the first fish turning, which is an informative source about all other fish turning afterwards.
For the last fish that turns the peak is negative, since this fish is misinformative about all other fish that have already turned.
The second, third and fourth fish present both a negative and a positive peak.
The intensity of the negative peaks increases from the second fish to the fourth, while the intensity of the positive peak decreases.

In general, the source fish is informative about all destination fish turning after it and misinformative about any destination fish turning before it.
This is because the prior turn of a source helps the observer to predict the later turn of the destination, whereas examining a source which has not turned yet itself is actively unhelpful (misinformative) in predicting the occurrence of such a turn.
This also explains why, for a source, the negative peaks come before positives.

The sequential cascade-like dynamics of information flow suggests that the strongest sources of predictive information transfer are fish that have already turned.
Moreover our analyses reveal that once a fish has performed a U-turn, its behaviour in general ceases to be predictable based on the behaviour of other fish that swim in opposite direction (in fact such fish would provide misinformative predictions).
This suggests an asymmetry of predictive information flows based on and about an individual fish during U-turns.

%%%%%%%%%%%%%%%%%%%%%%%%%%%%%
%%% Spatial motifs of information transfer %%%
%%%%%%%%%%%%%%%%%%%%%%%%%%%%%

\subsection*{Spatial motifs of information transfer}

\begin{figure}[t!]
\centering
\subfloat[]{\label{fig:roses-pos}\includegraphics[width=0.45\columnwidth]{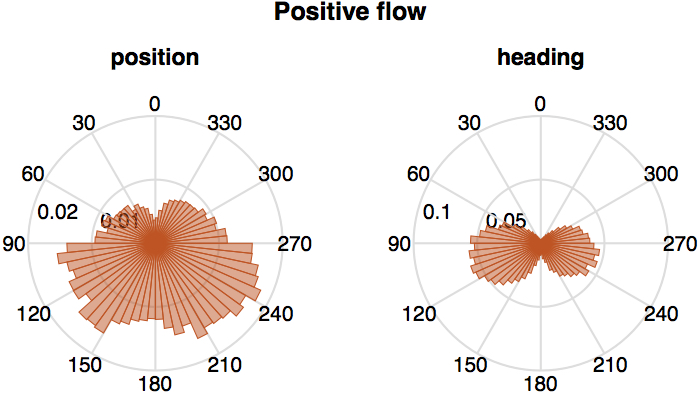}}\hspace{5mm}
\subfloat[]{\label{fig:roses-neg}\includegraphics[width=0.45\columnwidth]{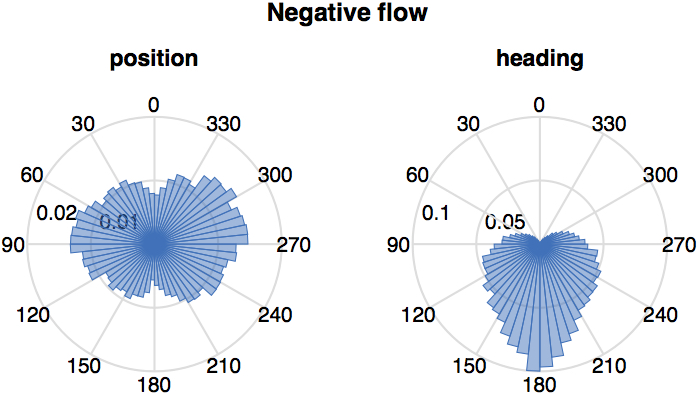}}
\caption{
Spatio-informational motifs.
Each diagram is a circle centred on a source fish with zero heading, providing a reference.
In each diagram space is divided into 60 angular sectors measuring 6\degree.
Within each circle we group all pairwise samples from all 455 U-turns such that the source fish is placed in the centre and the destination fish is placed within the circle in one of the sectors.
The left circles in Figures \ref{fig:roses-pos} and \ref{fig:roses-neg} aggregate the relative positions of destination fish, while the right circles aggregate the relative headings of destination fish.
The value of each radial sector (for both position and heading) represents the average of the corresponding values of either positive (Figure \ref{fig:roses-pos}) of negative (Figure \ref{fig:roses-neg}) transfer entropy.
For example, the value in each sector of the left diagram of \ref{fig:roses-pos} represents the average positive transfer entropy for a destination fish, given it has relative position in that sector with respect to the source fish: all positive values of transfer entropy corresponding to each sector are summed and divided by the total number of values corresponding to that sector.
The value in each sector of the right diagram of \ref{fig:roses-pos} represents the average positive transfer entropy for a destination fish, given that its heading diverges from the one of the source by an angle in that sector.
Figure \ref{fig:roses-neg} mirrors Figure \ref{fig:roses-pos} this negative transfer entropy.
The source fish data are taken from the time points corresponding to the time delay $v$ with respect to the source.
}
\label{fig:roses}
\end{figure}

It is reasonable to assume that predictive information transfer in a school of fish results from spatial interactions among individuals.
We investigated the role of pairwise spatial interactions in carrying the positive and negative information flows that we detected in the previous section, looking for spatial patterns of information and misinformation transfer.

In particular we established the statistics of the relative position and heading of the destination fish relative to the source fish, at times when the transfer entropy from the source to the destination is more intense.
For this purpose we used radial diagrams (see Figure \ref{fig:roses}) representing the relative data in terms transfer entropy, focusing separately on their positive (informative) and negative (misinformative) values.
In each diagram we aggregate data from all 455 U-turns and all pairs.
The diagrams show clear spatial patterns coupled with the transfer entropy, which we call spatio-informational motifs.

We see that positive information transfer is on average more intense from source fish to: a. other fish positioned behind them (Figure \ref{fig:roses-pos}, left), and b. to fish with headings closer to perpendicular rather than parallel to them (Figure \ref{fig:roses-pos}, right).
We know from Figures \ref{fig:trajectories} and \ref{fig:cascade} that positive transfer entropy is detected from source fish that have already turned to destination fish that are turning.
Thus, Figure \ref{fig:roses-pos} suggests that a source is more informative about destination fish that are left behind it after a turn, most intensely when the destination fish are executing their own turning manoeuvre to follow the source.
Directional relationships from individuals in front towards others that follow were observed in previous works on birds~\cite{nagy2010hierarchical}, bats~\cite{orange2015transfer} and fish~\cite{katz2011inferring, herbert2011inferring, rosenthal2015revealing}.

For negative information transfer (Figure \ref{fig:roses-neg}) we see a different spatio-informational motif.
Negative information transfer is on average more intense to fish generally positioned at the side and with opposite heading.
This aligns with Figures \ref{fig:trajectories} and \ref{fig:cascade} in that negative transfer entropy typically flows from fish that have not turned yet to those which are turning.

In summary, transfer entropy has a clear spatial signature, showing that the spatiotemporal dependencies in the studied school of fish are not random but reflect specific interactions.

%%%%%%%%%%%%%%%%%%%%%%%%%%%%%%%%%%%%%%%%%%%%%%%%%%%%%%%%%%%%%%%%%%%%%%
%%%%%%%%%%%%%%%%%%%%%%%%%%%%%% DISCUSSION %%%%%%%%%%%%%%%%%%%%%%%%%%%%%%%
%%%%%%%%%%%%%%%%%%%%%%%%%%%%%%%%%%%%%%%%%%%%%%%%%%%%%%%%%%%%%%%%%%%%%%

\section*{Discussion}

Information transfer within animal groups during collective motion is hard to quantify because of implicit and distributed communication channels with delayed and long-ranged effects, selective attention~\cite{riley1976multidimensional} and other species-specific cognitive processes.
Here we presented a rigorous framework for detecting and measuring predictive information flows during collective motion, by attending to the dynamic statistical dependence of directional changes in destination fish on relative heading of sources.
This predictive information flow should be interpreted as a change (gain or loss) in predictability obtained by an observer.

We studied \textit{Hemigrammus rhodostomus} fish placed in a ring-shaped tank which effectively only allowed the fish to move straight ahead or turn back to perform a U-turn.
The individual trajectories of the fish were recorded for hundreds of collective U-turns,
enabling us to perform a statistically significant information-theoretical analysis for multiple pairs of source and destination fish.

Transfer entropy was used in detecting pairwise time delayed dependencies within the school.
By observing the local dynamics of this measure, we demonstrated that predictive information flows intensify during collective direction changes --- i.e.~the U-turns ---
a hypothesis that until now was not verified in a real biological system.
Furthermore, we identified two distinct predictive information flows within the school: an informative flow based on fish that have already preformed the U-turn about fish that are turning, and a misinformative flow based on fish that have not preformed the U-turn yet about the fish that are turning.

We also explored the role of spatial dynamics in generating the influential interactions that carry the information flows, another well-known problem.
In doing so, we mapped the detected values of transfer entropy against fish relative position and heading, identifying clear spatio-informational motifs.
Importantly, the positive and negative predictive information flows were shown to be associated with specific spatial signatures of source and destination fish.
For example, positive information flow is detected when the source fish is in front of the destination, similarly to what was already observed in previous works on animals~\cite{nagy2010hierarchical, katz2011inferring, herbert2011inferring, rosenthal2015revealing, orange2015transfer}.

Local transfer entropy as it was applied in this study reveals the dynamics of \emph{pairwise} information transfer.
It is well-known that multivariate extensions to the transfer entropy, e.g. conditioning on other information sources, can be useful in terms of eliminating redundant pairwise relationships whilst also capturing higher-order relationships beyond pairwise (i.e. synergies) \cite{lizier2008local,lizier2010information,lizier2010differentiating,vak09,will11a}, and as such the identification of \emph{effective} neighbourhoods cannot be accurately inferred using pairwise relationships alone.
Improvements are possible by adapting algorithms for deciding when to include higher-order multivariate transfer entropies (and which variables to condition on), developed to study effective networks in brain imaging data sets~\cite{liz12c,faes11a,mar12c,stra12b}, to collective animal behaviour, as such methods can eliminate redundant connections and detect synergistic effects.
Whether or not such algorithms will prove useful for swarm dynamics is an open research question, with conflicting findings that first suggest that multiple fish interactions could be faithfully factorised into simply pair interactions in one species~\cite{gautrais2012deciphering} but conversely that this may not necessarily generalise~\cite{katz2011inferring}.

In any case, such adaptations to capture multivariate effects will be non-trivial, as it must handle the short-term and dynamic structure of interactions across the collective.
Early attempts have been made using (a similar measure to) conditional TE -- on average over time windows -- in collectives under such algorithms \cite{lord2016inference}, however it remains to be seen what such measures reveal about the collective dynamics on a local scale.

In summary, we have proposed a novel information-theoretic framework for studying the dynamics of information transfer in collective motion and applied it to a school of fish, without making any specific assumptions on fish behavioural traits and/or rules of interaction.
This framework can be easily applied to studies of other biological collective phenomena, such as swarming and flocking, artificial multi-agent systems and active matter in general.

%%%%%%%%%%%%%%%%%%%%%%%%%%%%%%%%%%%%%%%%%%%%%%%%%%%%%%%%%%%%%%%%%%%%%%
%%%%%%%%%%%%%%%%%%%%%%%%%%%%%%% METHODS %%%%%%%%%%%%%%%%%%%%%%%%%%%%%%%%
%%%%%%%%%%%%%%%%%%%%%%%%%%%%%%%%%%%%%%%%%%%%%%%%%%%%%%%%%%%%%%%%%%%%%%

\section*{Methods}

%%%%%%%%%%%%%%%%%
%%% Ethics statement %%%
%%%%%%%%%%%%%%%%%

\subsection*{Ethics statement}
All experiments have been approved by the Ethics Committee for Animal Experimentation of the Toulouse Research Federation in Biology N1 and comply with the European legislation for animal welfare.

%%%%%%%%%%%%%%%%%%%%%
%%% Experimental procedures %%%
%%%%%%%%%%%%%%%%%%%%%

\subsection*{Experimental procedures} 

70 \textit{Hemigrammus rhodostomus} (rummy-nose tetras) were purchased from Amazonie Lab{\`e}ge (\url{http://www.amazonie.com}) in Toulouse, France.
Fish were kept in 150 L aquariums on a 12:12 hour, dark:light photoperiod, at 27.7\degree C ($\pm0.5$\degree C) and were fed \textit{ad libitum} with fish flakes. Body lengths of the fish used in these experiments were on average 31 mm ($\pm$ 2.5 mm).

The experimental tank measured $120 \times 120$ cm, was made of glass and set on top of a box to isolate fish from vibrations.
The setup, placed in a chamber made by four opaque white curtains, was surrounded by four LED light panels giving an isotropic lighting.
A ring-shaped tank made from two tanks (an outer wall of radius 35 cm and an inner wall, a cone of radius 25 cm at the bottom, both shaping a corridor of 10 cm) was set inside the experimental tank filled with 7 cm of water of controlled quality (50\% of water purified by reverse osmosis and 50\% of water treated by activated carbon) heated at 28.1\degree C ($\pm 0.7$\degree C).
The conic shape of the inner wall has been chosen to avoid the occlusion on videos of fish swimming too close to the inner wall that would occur with straight walls.

Five fish were randomly sampled from their breeding tank for a trial.
Fish were ensured to be used only in one experiment per day at most.
Fish were let for 10 minutes to habituate before the start of the trial.
A trial consisted in one hour of fish swimming freely (i.e. without any external perturbation).

%%%%%%%%%%%%%%%%%%%%%%%%%%%
%%% Data extraction and pre-processing %%%
%%%%%%%%%%%%%%%%%%%%%%%%%%%

\subsection*{Data extraction and pre-processing}

Fish trajectories were recorded by a Sony HandyCam HD camera filming from above the setup at 50Hz (50 frames per second) in HDTV resolution (1920$\times$1080p).
Videos were converted from MTS to AVI files with the command-line tool FFmpeg 2.4.3.
Positions of fish on each frame were tracked with the tracking software idTracker 2.1~\cite{perez2014idtracker}.

When possible, missing positions of fish have been manually corrected, only during the collective U-turn events detected by the sign changes of polarisation of the fish groups.
The corrections have involved manual tracking of fish misidentified by idTracker as well as interpolation or merging of positions in the cases where only one fish was detected instead of several because they were swimming too close from each others for a long time.
All sequences less or equal than 50 consecutive missing positions were interpolated.
Larger sequence of missing values have been checked by eye to check whether interpolating was reasonable or not --- if not, merging positions with closest neighbors was considered.

Time series of positions have been converted from pixels to meters and the origin of the coordinate system $\mathcal{O}(0, 0)$ has been set to the centre of the ring-shaped tank.
The resulting data set contains $\num{9273720}$ data points ($\num{1854744}$ for each fish) including all the ten trials.
Velocity was numerically derived from position using the symmetric difference quotient two-point estimation~\cite{larson1983symmetric}.
Heading was then computed as the four-quadrant inverse tangent of velocity and used to compute transfer entropy.

%%%%%%%%%%%%%
%%% Polarisation %%%
%%%%%%%%%%%%%

\subsection*{Polarisation}

The polarisation is used to represent the orientation of a fish or of the whole school around the tank, which can be clockwise or anti-clockwise.
Let $Z$ and $\dot{Z}$ be the two-dimensional position and normalised velocity of a fish, defined as Cartesian vectors with the centre of the tank being the origin --- in case of the whole school, $Z$ and $\dot{Z}$ are averaged over all fish.
The fish direction along an ideal circular clockwise rotation is described by a unit vector $z=\frac{\omega\times Z}{{|\omega\times Z|}}$, where $\omega$ is a vector orthogonal to plane of the rotation, chosen using the left-hand rule.

The polarisation is defined as $\dot{Z}\cdot z$, so that it is positive when the fish is swimming clockwise and negative when it is swimming anti-clockwise.
Also, the better $\dot{Z}$ is aligned with $z$ or $-z$, the higher is the intensity of the polarisation.
On the contrary, as $\dot{Z}$ deviates from $z$ or $-z$, the polarisation decreases and eventually reaches zero when $\dot{Z}$ and $z$ are orthogonal.
As a consequence, during a U-turn the intensity of the polarisation decreases and becomes zero at least once, before it increases again with the opposite sign.

%%%%%%%%%%%%%%%%%%%%%%%%%%%
%%% Local information dynamic measures %%%
%%%%%%%%%%%%%%%%%%%%%%%%%%%

\subsection*{Local transfer entropy}

Transfer entropy~\cite{schreiber2000measuring} is defined in terms of Shannon entropy, a fundamental measure in Information Theory~\cite{cover91} that quantifies the uncertainty of random variables.
Shannon entropy of a random variable $X$ is $H(X)=-\sum_{x\in X}p(x)\log_2p(x)$, where $p(x)$ is the probability of a specific instance $x$ of $X$.
$H(X)$ can be interpreted as the minimal expected number of bits required to encode a value of $X$ without losing information.
The joint Shannon entropy between two random variables $X$ and $Y$ is $H(X,Y)=-\sum_{x\in X}\sum_{y\in Y}p(x,y)\log_2p(x,y)$, where $p(x,y)$ is the joint probability of instances $x$ of $X$ and $y$ of $Y$. 
This quantity allows the definition of conditional Shannon entropy as $H(X|Y)=H(X,Y)-H(X)$, which represents the uncertainty of $X$ knowing $Y$.

In this study we are interested in local (or pointwise) transfer entropy~\cite{fano61,liz14b} for specific instances of time-series processes of fish motion, which allows us to reconstruct the dynamics of information flows over time.
Shannon information content of an instance $x_n$ of process $X$ at time $n$ is defined as $h(x_n)=-\log_2 p(x_n)$.
The quantity $h(x_n)$ is the information content attributed to the specific instance $x_n$, or the information required to encode or predict that specific value.
Conditional Shannon information content of an instance $x_n$ of process $X$ given an instance $y_n$ of process $Y$ is defined as $h(x_n|y_n)=h(x_n,y_n)-h(x_n)$.

Local transfer entropy is defined as the information provided by the source $\mathbf{y_{n-v}} = \{y_{n-v}, y_{n-v-1},\allowbreak \ldots, y_{n-v-l+1}\}$, where $v$ is a time delay and $l$ is the history length, about the destination $x_n$ in the context of the past of the destination $\mathbf{x_{n-1}}=\{x_{n-1}, x_{n-2}, \ldots, x_{n-k}\}$, with a history length $k$:
\begin{equation}
\begin{aligned}
t_{y\to x}(n,v) &= h(x_n | \mathbf{x_{n-1}}) -h(x_n | \mathbf{x_{n-1}}, \mathbf{y_{n-v}})\\
&= \log_2\frac{ p( x_{n} | \mathbf{x_{n-1}} , \mathbf{y_{n-v}} ) }{ p( x_{n} | \mathbf{x_{n-1}} ) }.
\end{aligned}
\label{eq:teInMethods}
\end{equation}
Transfer entropy $T_{Y\to X}(v)$ is the average of the local transfer entropies $t_{y\to x}(n,v)$ over samples (or over $n$ under a stationary assumption).
The transfer entropy is asymmetric in $Y$ and $X$ and is also a dynamic measure (rather than a static measure of correlations) since it measures information in state transitions of the destination.

In order to compute transfer entropy here, the source variable $Y$ and destination variable $X$ are defined in terms of the fish heading.
Specifically, $X$ is the first-order divided difference (Newton's difference quotient) of the destination fish heading, while $Y$ is the difference between the two fish headings at the same time.
Let $\Theta_S$ and $\Theta_D$ be respectively the heading time series of the source and the destination fish. We then construct variables $X$ and $Y$ as follows, for all time points $n$:
\begin{center}
\setlength\tabcolsep{0pt}
\begin{tabular}{m{0.55\columnwidth}m{0.45\columnwidth}}
\centering
\includegraphics[height=3.5cm]{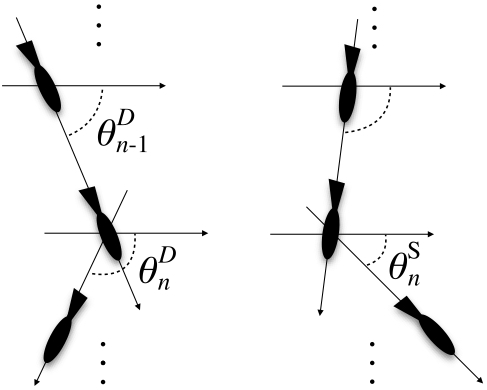} &
\begin{equation}
x_n = \Theta^D_n - \Theta^D_{n-1} \textrm{\ \ \ \ \ \ \ }
\label{eq:fish-x}
\end{equation}
\vspace{0.2cm}
\begin{equation}
y_n = \Theta^D_n - \Theta^S_n . \textrm{\ \ \ \ \ \ \ \ \ }
\label{eq:fish-y}
\end{equation}
\end{tabular}
\setlength\tabcolsep{6pt}
\end{center}
Thus, $y_n$ represents the relative heading of the destination fish with respect to the source fish, while $x_n$ represents the directional change of the destination fish.
The variables were so defined in order to capture directional changes of the destination fish in relation to its alignment with the source fish, which is considered an important component of movement updates in swarm models \cite{reynolds87}.

Given the definition of the variables \eqref{eq:fish-x} and \eqref{eq:fish-y}, we computed local transfer entropy $t_{y\to x}(n,v)$ using Equation \eqref{eq:teInMethods}, where $v$ was determined as described in section ``Parameters optimisation'' that follows.
The past state $\mathbf{x_{n-1}}$ of the destination in transfer entropy was defined as a vector of an embedding space of dimensionality $k$ and delay $\tau$, with $\mathbf{x_{n-1}} = \{x_{n-1-j\tau}\}$, for $j = \{0,1,\dots,k-1\}$.
Finding optimal values for $k$ and $\tau$ is also described in section ``Parameters optimisation''.
The state of the source process $\mathbf{y_{n-v}}$ was also defined as a vector of an embedding space whose the dimensionality $l$ and delay $\tau'$ were similarly optimised.
The local transfer entropy $t_{y\to x}(n,v)$ computed on these variables therefore tells us how much information ($l$ time steps of) the heading of the destination relative to the source adds to our knowledge of the directional change in the destination (some $v$ time steps later), in the context of $k$ past directional changes of the destination.
We note that while turning dynamics of the destination may contain more entropy (as rare events), there will only be higher transfer entropy at these events if the source fish is able to add to the prediction of such dynamics.

Computing transfer entropy requires knowledge of the probabilities of $x_n$ and $y_n$ defined in \eqref{eq:fish-x} and \eqref{eq:fish-y}.
These are not known a priori, but the measures can be estimated from the data samples using existing techniques.
In this study, this was accomplished assuming that the probability distribution function for the observations is a multivariate Gaussian distribution (making the transfer entropy proportional to the Granger causality~\cite{barnett2009granger}), using the JIDT software implementation~\cite{lizier2014jidt}.

Also, we assume stationarity of behaviour and homogeneity across the fish, such that we can pool together all pairwise samples from all time steps, for all trials, maximising the number of samples available for the calculation of each measure.
For performance efficiency, we make calculations of the local measures using 10 separate sub-sampled sets (sub-sampled evenly across the trials), then recombine into a single resultant information-theoretic data set.

%%%%%%%%%%%%%%%%%%%%
%%% Parameters optimisation %%%
%%%%%%%%%%%%%%%%%%%%

\subsection*{Parameter optimisation}

The embedding dimensionality and delay for the source and the past state of the destination need to be appropriately chosen in order to optimise the quality of transfer entropy.
The combination $(k,\tau)$ for the past state of the destination, as well as the combination $(l,\tau')$ for the source, have been optimised separately by minimising the global self-prediction error, as described in~\cite{PhysRevE.65.056201,wib14c}.
In the case of Markov processes, the optimal dimensionality of the embedding is the order of the process.
Lower dimensions do not provide the same amount of predictive information, while higher dimensions add redundancy that weaken the prediction.
For non-Markov processes, the algorithm selects the highest dimensionality found to contribute to self-prediction of the destination whilst still being supported by the finite amount of data that we have.
Values of the dimensionality between 1 and 10 have been explored in combination with values of the delay between 1 and 5. The optimal combinations were found to be the same for both the source and the past of the destination: $k=l=3$, $\tau=\tau'=1$.

\begin{figure}[t]
\centering
\includegraphics[width=0.55\columnwidth]{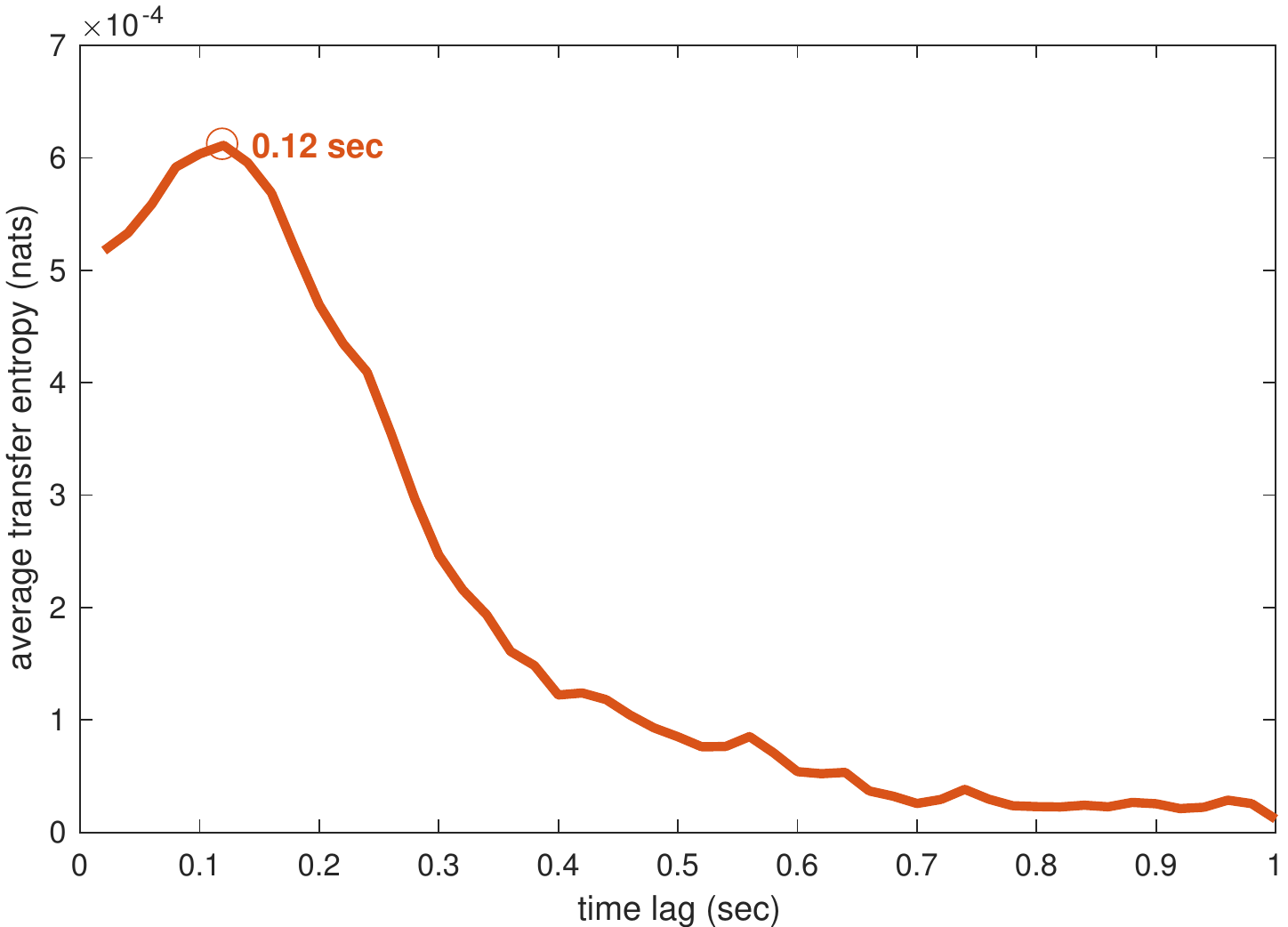}
\caption{
Time lag optimisation. The red line represents the average transfer entropy (with $k=l=3$, $\tau=\tau'=1$) over all samples, as a function of the time delay between the source variable and the destination variable, for time delays between 0.02 to 1 seconds (1 to 50 time cycles).
}
\label{fig:lag-opt}
\end{figure}

The lag $v$ was also optimised.
This was done by maximising the average transfer entropy (after the optimisation of $k$, $\tau$, $l$ and $\tau'$) as per~\cite{wib13a}, over lags between 0.02 and 1 second, at time steps of 0.02 seconds.
The average transfer entropy was observed to grow and reach a local maximum at $v=6$ ($0.12$ seconds), and then decrease for higher values (see Figure \ref{fig:lag-opt}).
This result might have a biological interpretation: it is plausible for a fish to have a minimum reaction time, which delays the response to behaviour of other fish.

%%%%%%%%%%%%%%%%%%%%%%%%%%%%%%%%%%%%%%%%%%
%%% Statistical significance of estimates of local transfer entropy %%%
%%%%%%%%%%%%%%%%%%%%%%%%%%%%%%%%%%%%%%%%%%

\subsection*{Statistical significance of estimates of local transfer entropy}

Theoretically, transfer entropy between two independent variables is zero.
However, a non-zero bias (and a variance of estimates around that bias) is likely to be observed when, as in this study, transfer entropy is numerically estimated from a finite number of samples.
This leads to the problem of determining whether a non-zero estimated value represents a real relationship between two variables, or is otherwise not statistically significant~\cite{wib14c}.

There are known statistical significance tests for the average transfer entropy~\cite{vic11a,liz11a,lizier2014jidt}, involving comparing the measured value to a null hypothesis that there was no (directed) relationship between the variables.
For an average transfer entropy estimated from $N$ samples, one surrogate measurement is constructed by resampling the corresponding $\mathbf{y_{n-v}}$ for each of the $N$ samples of $\{x_n , \mathbf{x_{n-1}} \}$ and then computing the average transfer entropy over these new surrogate samples.
This process retains $p(x_n | \mathbf{x_{n-1}})$ and $p(\mathbf{y_{n-v}})$, but not $p(x_n | \mathbf{y_{n-v}}, \mathbf{x_{n-1}})$.
Many surrogate measurements are repeated so as to construct a surrogate distribution under this null hypothesis of no directed relationship, and the transfer entropy estimate can then be compared in a statistical test against this distribution.
For the average transfer entropy measured via the linear-Gaussian estimator, it is known that analytically the surrogates (in nats, and multiplied by $2 \times N$) asymptotically follow a $\chi^2$ distribution with $l$ degrees of freedom \cite{gew82,barn12a}.
We use this distribution to confirm that the transfer entropy at the selected lag of 0.12 seconds (and indeed all lags tested) is statistically significant compared to the null distribution (at $p < 0.05$ plus a Bonferroni correction for the multiple comparisons across the 50 candidate lags).

Next, we introduce an extension of these methods in order to assess the statistical significance of the \emph{local} values.
This simply involves constructing surrogate transfer entropy measurements as before, however this time retaining the local values within those surrogate measurements and building a distribution of those surrogates.
Measured local values are then statistically tested against this null distribution of local surrogates to assess their statistical significance.

We generated ten times as many surrogate local values as the number of actual local estimates, with a total of approximately $371$ million local surrogates.
This large set of surrogate local values was used to estimate $p$-values of actual local values of the transfer entropy.
If $p$-value is sufficiently small, then the test fails and the value of the transfer entropy is considered significant (the value represents an actual relationship).
The Benjamini-Hochberg~\cite{10.2307/2346101} procedure was used to select the $p$-value cutoff whilst controlling for the false discovery rate under ($N$) multiple comparisons.

% ACKNOWLEDGEMENTS
\section*{Acknowledgements}
E.C. was supported by the University of Sydney's ``Postgraduate Scholarship in the field of Complex Systems'' from Faculty of Engineering \& IT and by a CSIRO top-up scholarship.
L.J. was supported by a grant from the China Scholarship Council (CSC NO.201506040167).
V.L. was supported by a doctoral fellowship from the scientific council of the University Paul Sabatier.
This study was supported by grants from the Centre National de la Recherche Scientifique and University Paul Sabatier (project Dynabanc).
J.L. was supported through the Australian Research Council DECRA grant DE160100630.
The University of Sydney HPC service provided computational resources that have contributed to the research results reported within this paper.

% AUTHORS' CONTRIBUTIONS
\section*{Author contributions statement}
G.T. designed research; V.L., P.T. and G.T. performed research; V.L., L.J., P.T., R.W. and G.T. analysed data.
E.C., J.L., R.W. and M.P. developed information dynamics methods, performed information-theoretic analysis, and identified information flows and motifs.
E.C. designed, developed and run software for the information-theoretic analysis.
G.T., J.L., E.C. and M.P. conceived and analysed information cascade.
E.C., J.L and M.P. wrote the paper.
G.T. and V.L. edited the manuscript and contributed to the writing.

% BIBLIOGRAPHY
\bibliographystyle{plain}
\bibliography{paper}

%%%%%%%%%%%%%%%%%%%%%%%%%%%%%%%%%%%%%%%%%%%%%%%%%%%%%%%%%%%%%%%%%%%%%%
%%%%%%%%%%%%%%%%%%%%%% SUPPLEMENTARY INFORMATION %%%%%%%%%%%%%%%%%%%%%%%%%%%
%%%%%%%%%%%%%%%%%%%%%%%%%%%%%%%%%%%%%%%%%%%%%%%%%%%%%%%%%%%%%%%%%%%%%%

\newpage

\setcounter{page}{1}

\begin{center}
\noindent\LARGE{Informative and misinformative interactions in a school of fish}\\
\ \\
\noindent\large{Emanuele Crosato, Li Jiang, Valentin Lecheval, Joseph T. Lizier,\\X. Rosalind Wang,
Pierre Tichit, Guy Theraulaz, Mikhail Prokopenko}\\
\end{center}
\ \\

\section*{Supplementary information}

% SUPPLEMENTARY FIGURE 1
\begin{figure}[h]
\captionsetup{labelformat=empty}
\centering
\includegraphics[width=\textwidth]{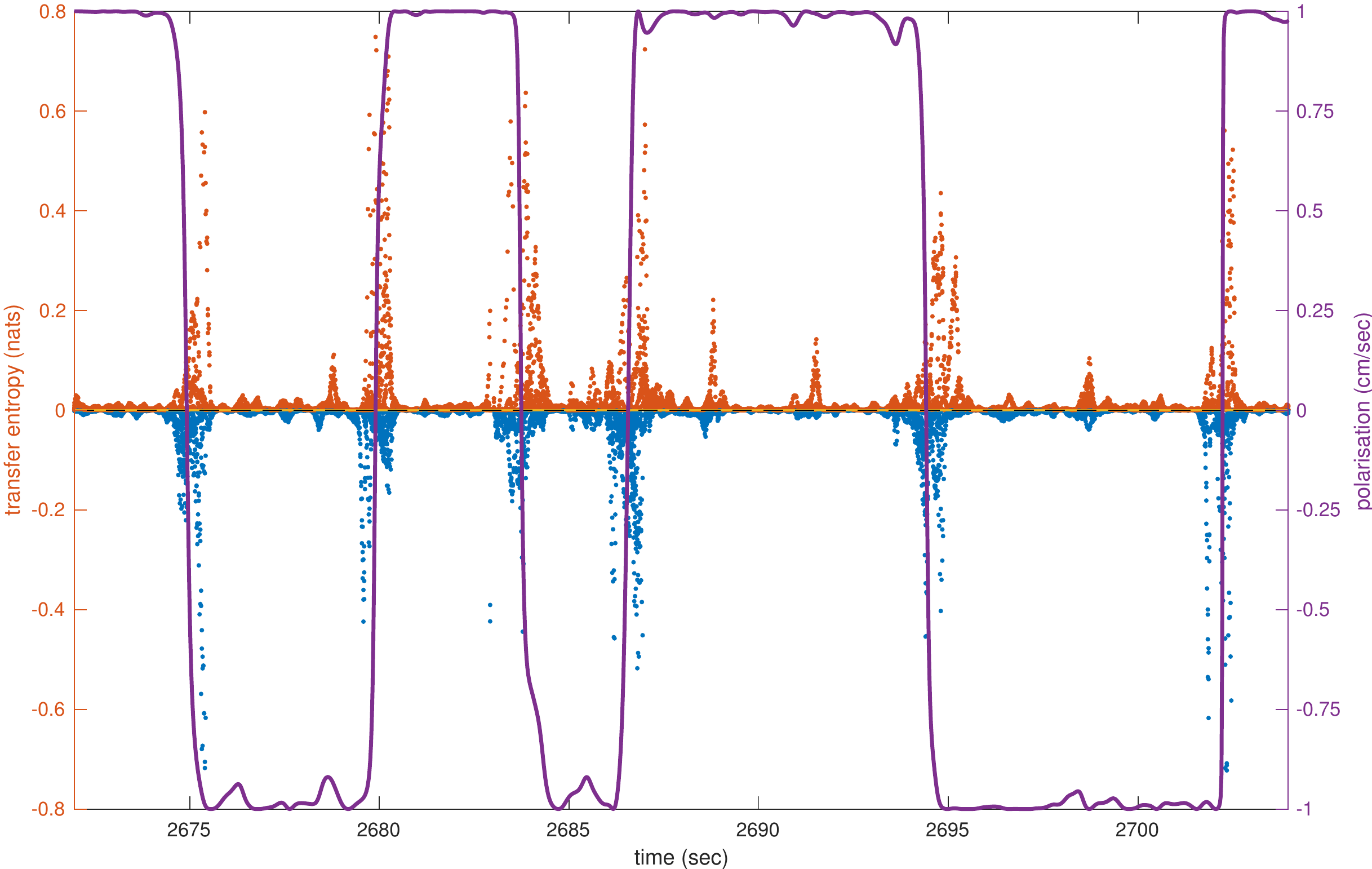}
\caption{
Supplementary Figure S1.
Transfer entropy within the school during several U-turns.
The figure plots the school's polarisation during a U-turn and the detected transfer entropy over a time interval of approximately 35 seconds.
The purple line represents the school's polarisation, while dots represent local values of transfer entropy between all directed pairs of fish: red dots represent positive transfer entropy and blue dots represent negative transfer entropy.
Time is discretised in steps of length 0.02 seconds and for each time step 20 points of these local measures are plotted, for the 20 directed pairs formed out of 5 fish.
}
\end{figure}

% SUPPLEMENTARY VIDEO 1
\begin{figure}[h]
\captionsetup{labelformat=empty}
\centering
\href{https://www.dropbox.com/s/otdjaupw4m75n2l/Supplementary\%20Video\%20S1.mp4?dl=0}{\includegraphics[width=\textwidth]{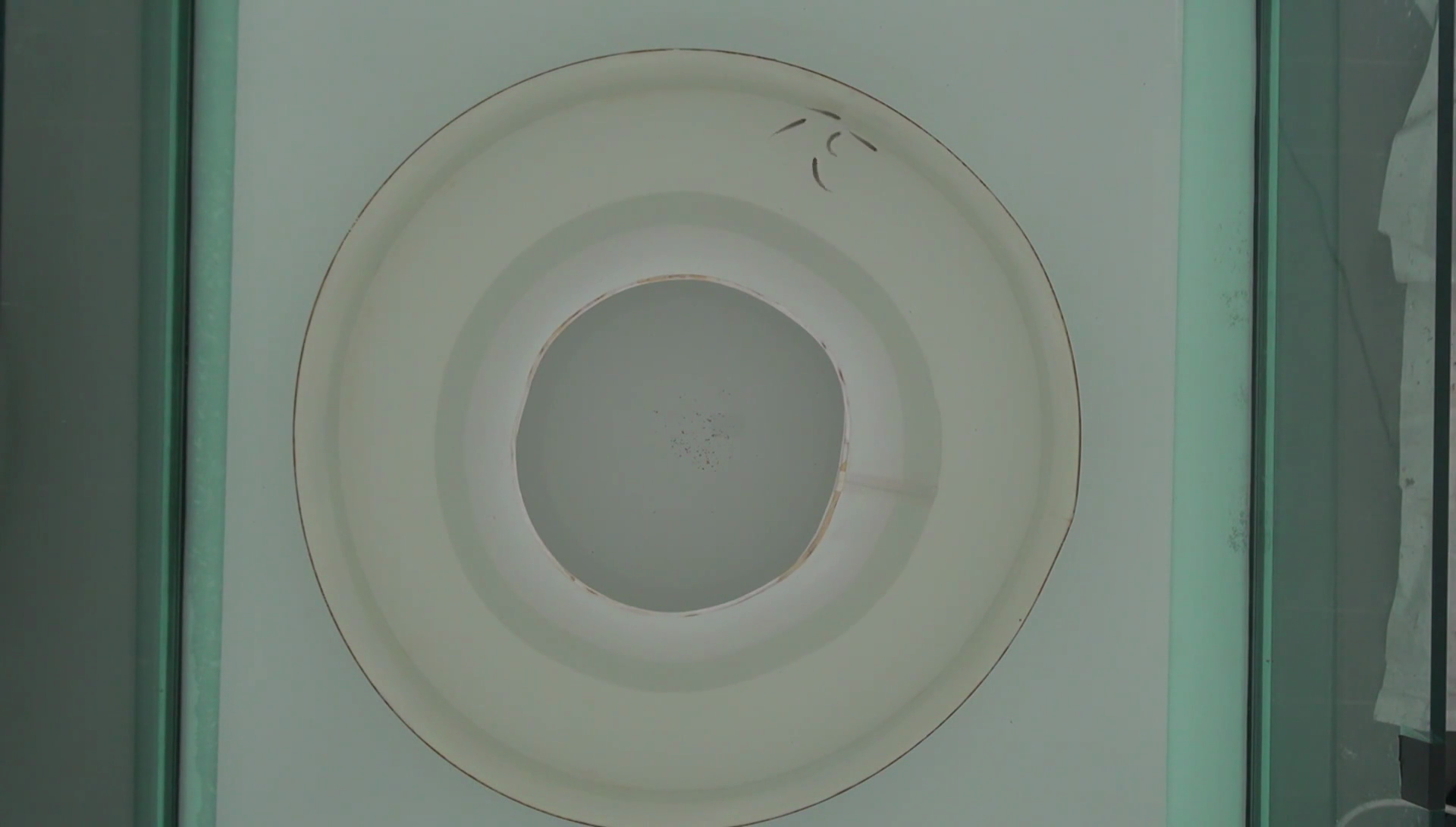}}
\caption{Supplementary Video S1.
Fish undergoing the representative U-turn (click on the image to open the video).
The movie shows five \textit{Hemigrammus rhodostomus} swimming in the ring-shaped tank for approximately 6 seconds, during which they undergo the U-turn presented in the main article.
}
\end{figure}

% SUPPLEMENTARY VIDEO 2
\begin{figure}[h]
\captionsetup{labelformat=empty}
\centering
\href{https://www.dropbox.com/s/up36icwkf9dmxyt/Supplementary\%20Video\%20S2.mp4?dl=0}{\includegraphics[width=\textwidth]{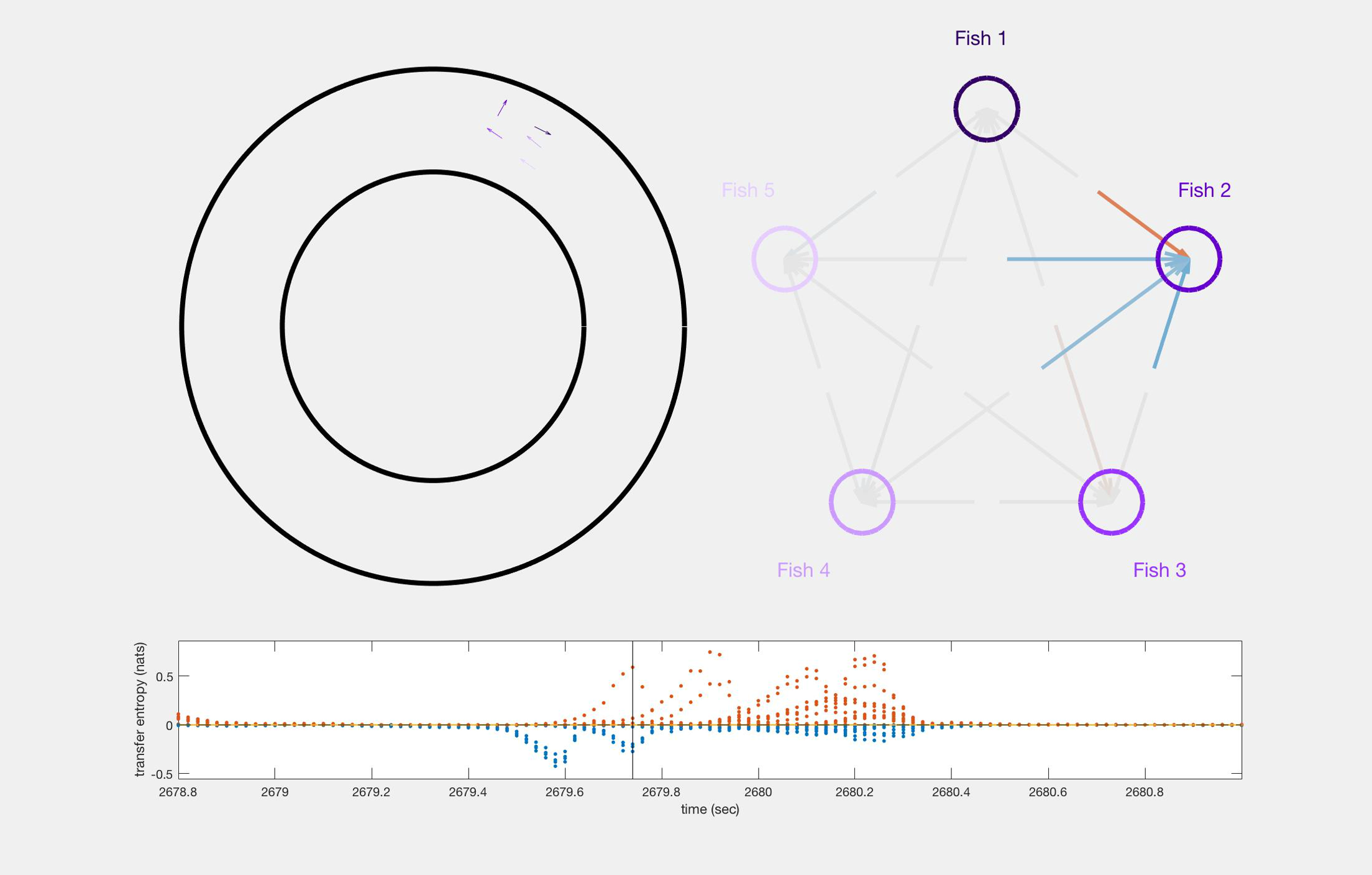}}
\caption{Supplementary Video S2 (click on the image to open the video).
Animation of the representative U-turn showing transfer entropy dynamics.
The movie shows an animation of the representative U-turn over a time interval of approximately 2 seconds.
On the top-left is the ring-shaped tank with the five fish, represented by arrows of different shades of purple.
On the bottom is the transfer entropy between any directed pair of fish over the time interval: red dots represent positive transfer entropy and blued dots represent negative transfer entropy.
Time is discretised in steps of length 0.02 seconds and for each time step 20 points of transfer entropy are plotted, for the 20 directed pairs that can be formed out of 5 fish.
On the top-right is the network of transient neighbours changing over time.
Each node represents a fish and each directed edge entering a node indicates the transfer entropy to that fish from the other four (the source fish is easily identifiable from the angle of the edges).
The colour of the edges changes during the U-turn: strong red indicates intense positive transfer entropy; strong blue indicates intense negative transfer entropy; intermediate grey indicates that transfer entropy is close to zero.
}
\end{figure}

\end{document}